\documentclass[journal]{IEEEtran}
\IEEEoverridecommandlockouts
\usepackage{booktabs}
\usepackage{siunitx}
\usepackage{cite}
\usepackage{amsmath,amssymb,amsfonts}
\usepackage{graphics, framed}
\usepackage{graphicx, enumerate}
\usepackage{epsfig}
\usepackage{epstopdf}
\usepackage{url}
\usepackage{multimedia}
\usepackage{paralist}
\usepackage[printonlyused]{acronym}
\usepackage{units}
\usepackage{algorithmic}
\usepackage{textcomp}
\usepackage{float}
\usepackage{tabularx}
\usepackage{balance}
\usepackage{subfigure}
\usepackage{xcolor}
\def\BibTeX{{\rm B\kern-.05em{\sc i\kern-.025em b}\kern-.08em
T\kern-.1667em\lower.7ex\hbox{E}\kern-.125emX}}
\usepackage{mathtools}
\usepackage{cuted} 
\usepackage[export]{adjustbox}

\usepackage{amsmath}
\usepackage{amssymb}
\usepackage{dblfloatfix}
\usepackage{lipsum}
\usepackage{color,soul}
\usepackage{multirow}
\usepackage[overload]{textcase}
\newcommand{\FGR}[1]{Fig.~\ref{#1}}

\newcommand{\SEC}[1]{Section~\ref{#1}}
\newcommand{\TAB}[1]{Table~\ref{#1}}

\acrodef{5G}[5G]{5\textsuperscript{th}-Generation}
\acrodef{BW}[BW]{bandwidth}
\acrodef{BER}[BER]{bit error rate}
\acrodef{BPSK}[BPSK]{binary phase-shift keying}
\acrodef{CW}[CW]{continuous wave}
\acrodef{CSI}[CSI]{channel state information}
\acrodef{D2D}[D2D]{device-to-device}
\acrodef{dB}[dB]{decibel}
\acrodef{dBi}[dBi]{decibel isotropic}
\acrodef{dBm}[dBm]{decibel over a milliwatt}
\acrodef{DSN}[DSN]{deep space network}
\acrodef{DTN}[DTN]{delay tolerant network}
\acrodef{Gbps}[Gbps]{gigabit per second}
\acrodef{GHz}[GHz]{gigahertz}
\acrodef{GAT}[GAT]{graph attention network}
\acrodef{GAL}[GAL]{graph attention layer}
\acrodef{THz}[THz]{Terahertz}
\acrodef{ISL}[ISL]{inter-satellite link}
\acrodef{RIS}[RIS]{reconfigurable intelligent surface}
\acrodef{TDMA}[TDMA]{time division multiple access}
\acrodef{GM}[GM]{Gamma mixture}
\acrodef{PSK}[PSK]{phase shift keying}
\acrodef{QAM}[QAM]{quadrature amplitude modulation}
\acrodef{AWGN}[AWGN]{additive white Gaussian noise}
\acrodef{SNR}[SNR]{signal-to-noise ratio}
\acrodef{AF}[AF]{amplitude-and-forward}
\acrodef{MIMO}[MIMO]{multiple-input multiple-output}
\acrodef{mMIMO}[mMIMO]{massive-multiple-input multiple-output}
\acrodef{SDN}[SDN]{Software-defined network}
\acrodef{SON}[SON]{self-organizing network}
\acrodef{HetNet}[HetNet]{heterogeneous network}
\acrodef{FSO}[FSO]{free-space optics}
\acrodef{UM-MIMO}[UM-MIMO]{ultra-massive-MIMO}
\acrodef{AP}[AP]{access point}
\acrodef{UE}[UE]{user equipment}
\acrodef{NTN}[NTN]{non-terrestrial network}
\acrodef{UAV}[UAV]{unmanned aerial vehicle}
\acrodef{HAPS}[HAPS]{high-altitude platform station}
\acrodef{LEO}[LEO]{low-Earth orbit}
\acrodef{BAN}[BAN]{body area network}
\acrodef{WLAN}[WLAN]{wireless local area network}
\acrodef{QoS}[QoS]{quality of service}
\acrodef{TCS}[TCS]{thermal control system}
\acrodef{QCL}[QCL]{quantum cascade laser}
\acrodef{CMOS}[CMOS]{complementary metal-oxide semiconductor}
\acrodef{V-HetNet}[V-HetNet]{vertical heterogeneous network}
\acrodef{DL}[DL]{deep learning}
\acrodef{DRL}[DRL]{deep reinforcement learning}
\acrodef{EIRP}[EIRP]{effective isotropic radiated power}
\acrodef{FDTD}[FDTD]{Finite-difference time-domain}
\acrodef{FEM}[FEM]{finite element method}
\acrodef{MoM}[MoM]{method of moments}
\acrodef{VNA}[VNA]{vector network analyzer}
\acrodef{CNN}[CNN]{convolutional neural network}
\acrodef{CIR}[CIR]{channel impulse response}
\acrodef{CTF}[CTF]{channel transfer function}
\acrodef{DPM}[DPM]{Dirichlet process mixture}
\acrodef{TOA}[TOA]{time of arrival}
\acrodef{GMM}[GMM]{Gaussian mixture model}
\acrodef{IoT}[IoT]{Internet of things}
\acrodef{TDD}[TDD]{time-division duplex}
\acrodef{MLE}[MLE]{maximum likelihood estimation}
\acrodef{LOS}[LOS]{line-of-sight}
\acrodef{NLOS}[NLOS]{non-line-of-sight}
\acrodef{SG}[SG]{signal generator}
\acrodef{SEP}[SEP]{Sun-Earth-probe}
\acrodef{FDSOI}[FDSOI]{fully depleted silicon on insulator}
\acrodef{OpEx}[OpEx]{operational expenditures}
\acrodef{TCO}[TCO]{total cost of ownership}
\acrodef{CapEx}[CapEx]{capital expenditures}
\acrodef{MAC}[MAC]{medium access control}
\acrodef{GEO}[GEO]{geostationary orbit}
\acrodef{SWaP}[SWaP]{size, weight, and power}
\acrodef{CSI}[CSI]{channel state information}
\acrodef{PN}[PN]{pseudo-noise}
\acrodef{NMSE}[NMSE]{normalized mean square error}
\acrodef{MSE}[MSE]{mean square error}
\acrodef{LS}[LS]{least square}
\acrodef{DtS}[DtS]{direct-to-satellite}

\ifCLASSINFOpdf
\else
\fi

\begin{document}
\title{Graph Attention Networks for Channel Estimation in RIS-assisted Satellite IoT Communications}

\author{K{\"{u}}r{\c{s}}at~Tekb{\i}y{\i}k,~\IEEEmembership{Graduate Student Member,~IEEE,} G{\"{u}}ne{\c{s}}~Karabulut~Kurt,~\IEEEmembership{Senior~Member,~IEEE,} Ali~R{\i}za~Ekti,~\IEEEmembership{Senior~Member,~IEEE,} Halim~Yanikomeroglu,~\IEEEmembership{Fellow,~IEEE}

\thanks{K. Tekb{\i}y{\i}k is with the Department of Electronics and Communications Engineering, {\.{I}}stanbul Technical University, {\.{I}}stanbul, Turkey, e-mail: tekbiyik@itu.edu.tr}
\thanks{G. Karabulut Kurt is with the Poly-Grames Research Center, Department of Electrical Engineering,  Polytechnique Montr\'eal, Montr\'eal, Canada, e-mail: gunes.kurt@polymtl.ca}

\thanks{A.R. Ekti is with the Grid Communications and Security Group, Electrification and Energy Infrastructure Division, Oak Ridge National Laboratory, Oak Ridge, TN, U.S.A., e-mail: ektia@ornl.gov. This manuscript has been authored in part by UT-Battelle, LLC, under contract DE-AC05-00OR22725 with the US Department of Energy (DOE). The publisher acknowledges the US government license to provide public access under the DOE Public Access Plan (http://energy.gov/downloads/doe-public-access-plan).}

\thanks{H. Yanikomeroglu is with the Department of Systems and Computer Engineering, Carleton University, Ottawa, Canada, e-mail: halim@sce.carleton.ca}
 
}

\IEEEoverridecommandlockouts 

\maketitle

\begin{abstract}
Direct-to-satellite (DtS) communication has gained importance recently to support globally connected \ac{IoT} networks. However, relatively long distances of densely deployed satellite networks around the Earth cause a high path loss. In addition, since high complexity operations such as beamforming, tracking and equalization have to be performed in \ac{IoT} devices partially, both the hardware complexity and the need for high-capacity batteries of \ac{IoT} devices increase. The \acp{RIS} have the potential to increase the energy-efficiency and to perform complex signal processing over the transmission environment instead of \ac{IoT} devices. But, \acp{RIS} need the information of the cascaded channel in order to change the phase of the incident signal. \textcolor{black}{This study evaluates the pilot signal as a graph and incorporates this information into the graph attention networks (GATs) to track the phase relation through pilot signaling. \textcolor{black}{The proposed} GAT-based channel estimation method examines the performance of the DtS IoT networks for different RIS configurations to solve the challenging channel estimation problem.} It is shown that the proposed GAT both demonstrates a higher performance with increased robustness under changing conditions and has lower computational complexity compared to conventional \acl{DL} methods. Moreover, \acl{BER} performance is investigated for RIS designs with discrete and non-uniform phase shifts under channel estimation based on the proposed method. One of the findings in this study is that the channel models of the operating environment and the performance of the channel estimation method must be considered during RIS design to exploit performance improvement as far as possible.
\end{abstract}

\IEEEpeerreviewmaketitle
\acresetall

\section{Introduction}\label{sec:intro}
\ac{IoT} networks are expected to grow with approximately $20\%$ in terms of compound annual growth~\cite{nassar2022deep}. In other words, more than 100 billion devices will be connected in massive ubiquitous networks~\cite{sanislav2021energy, akpakwu2017survey}. This growth brings a backhauling gap for the \textcolor{black}{omnipresently} connected massive number of \ac{IoT} devices. In this context, dense \ac{LEO} satellite deployments can be an enabler for global service of \ac{IoT} devices. Low-power \ac{LEO} satellites have been already in service~\cite{fraire2019direct}. However, a new paradigm, which is called \ac{DtS}, has recently emerged to connect \ac{IoT} devices directly to satellites without any gateways on the ground~\cite{fraire2020sparse}. 

\begin{figure*}[ht]
    \centering
    \includegraphics[width=\linewidth, page=1]{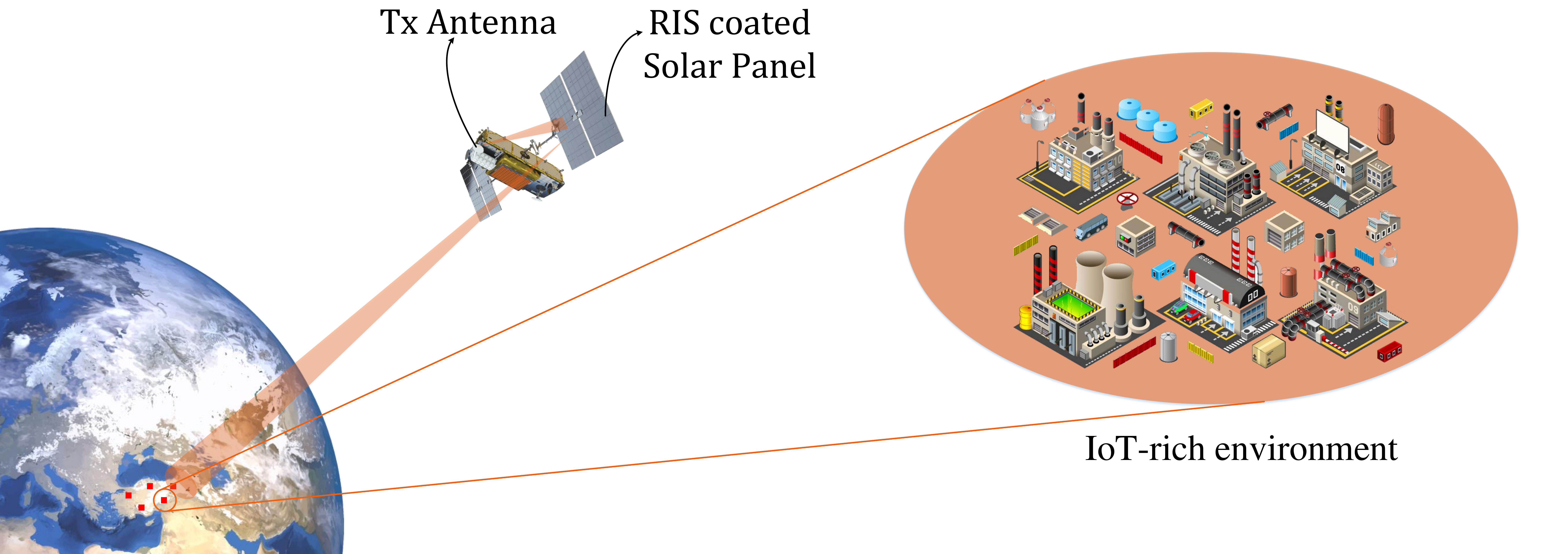}
    \caption{It is possible to enhance the QoS for satellite-IoT systems by utilizing RIS. Therefore, the required power can be reduced for the same data rate and error probability.}
    \label{fig:system_scheme}
\end{figure*}

Even though \ac{GEO} satellites have been proposed for narrowband-IoT applications~\cite{barbau2020nb, hofmann2019direct}, high delay and high path loss caused by the distance between \ac{GEO} satellites and the Earth reduce the energy-efficiency. On the other hand, \ac{LEO} satellite constellations become a prominent way to support global \ac{IoT} network with a reasonable path delays~\cite{cluzel20183gpp, qu2017leo}. Furthermore, the required transmit power for \ac{LEO} orbits is lower due to the relatively short distance to Earth. For \ac{LEO} satellite assisted \ac{IoT} communications, two methods are employed: indirect link and direct (i.e., \ac{DtS}) transmission links. Direct access is preferred due to the following reasons: the cost of the gateway infrastructure, temporary device deployments for specific environments, and operation ability after any disaster~\cite{fraire2019direct}. However, \ac{DtS} requires a steerable antenna at ground stations (i.e., \ac{IoT} devices) for tracking owing to the motion of \ac{LEO} satellites. Furthermore, a sophisticated transceiver is needed in \ac{IoT} devices to recover the received signal. Considering the hardware limitation and battery capacity for \ac{IoT} devices, these requirements cannot be met. Rather than an advanced transceiver in each \ac{IoT} device, it is possible to apply complex signal processing methods over the propagation medium~\cite{basar2019wireless}. 

\textcolor{black}{Smart artificial surfaces, referred to as \ac{RIS}, have been recently proposed to shift the processes on the receiver to \textcolor{black}{the} propagation medium by adjusting the incident wave phase~\cite{wu2019towards}. In other words, coding or complex processing is not needed by \acp{RIS}. The most appealing feature of \acp{RIS} is to comprise only passive elements with a single RF chain~\cite{tekbiyik2020reconfigurable}. \textcolor{black}{It is worth noting here that saturation on RISs is not expected due to the all-passive element structure of RISs although the distance between transmitter and RIS is close.} Thus, \textcolor{black}{the use of} RIS \textcolor{black}{fits} well \textcolor{black}{with} LEO satellites considering their \ac{SWaP} constraints~\cite{tekbiyik2020reconfigurable_sat}. The recent prototypes~\cite{dai2020reconfigurable, tang2020mimo, hu2020reconfigurable} demonstrate that it is possible to decrease the size and weight of the transceiver compared to the conventional \ac{MIMO} systems. Moreover, the battery life of \ac{IoT} devices is extended because of decreasing processing energy~\cite{huang2019reconfigurable}. It should be highlighted that \acp{RIS} can be a game-changer for \ac{DtS} \ac{IoT} systems since \acp{RIS} \textcolor{black}{can} reduce both the required transmit power and hardware complexity. Furthermore, our previous works~\cite{tekbiyik2020reconfigurable_sat, tekbiyik2021energy} present that \acp{RIS} can improve the system performance for LEO inter-satellite links and \textcolor{black}{achievable} data rate for satellite-\ac{IoT} systems, respectively. It is demonstrated that utilizing \acp{RIS} in \ac{DtS} system provides two-fold advantages. By focusing the beam towards the receiver, \ac{SNR} can be maximized while maintaining the same transmit power. The second advantage is the decrease in computational complexity of the transceiver by processing the transmitted signal on the propagation medium rather than \textcolor{black}{on} the receiver.}

\textcolor{black}{The current literature clearly shows that \acp{RIS} can provide many attractive features for \ac{DtS} \ac{IoT} systems. A few of these are the energy-efficiency, improved achievable rate, beamforming, and tracking. However, in order to take advantage of all these attractive features, \ac{CSI} must be obtained properly. The main challenge in channel estimation for \ac{RIS}-assisted communications is that the received signal includes cascaded channel coefficients. To cope up with the channel estimation problem, some methods have been proposed. The first methodology~\cite{elbir2020deep, mishra2019channel} utilizes a conventional view for channel estimation by activating a single \ac{RIS} element for each time instance during the training phase. In other words, the channel coefficients are individually estimated by switching \ac{RIS} elements on and off. Considering the switching time and total pilot overhead, it can be concluded that this method cannot be efficiently employed on \ac{RIS}-assisted \ac{DtS} system because of the longer training signal than channel coherence time~\cite{corazza1994statistical}. Concisely, acquired channel coefficients regarding \ac{RIS} elements by this method are not time-invariant. In~\cite{chen2019channel}, the compressive sensing is employed to decrease pilot training overhead. But, its iterative approach to \textcolor{black}{solving a} non-convex problem might require longer computation time for simple-hardware systems. It is known that \acp{RIS} comprise only passive reflecting/scattering elements; hence, they cannot acquire \ac{CSI} by themselves. A recent approach in~\cite{taha2019enabling} designs \ac{RIS} in which some elements are active to estimate channel coefficients at \acp{RIS}. Please note that increasing the number of active elements on \ac{RIS} means that problems related to \ac{SWaP} constraints will probably arise. Therefore, this method is not desired for satellite systems.}

\textcolor{black}{To address the challenges related to the channel estimation in \ac{RIS}-assisted wireless communications, we have recently proposed a channel estimator based on \ac{GAT}~\cite{tekbiyik2020channel}. It should be noted that to the best knowledge of the authors, there is no channel estimation method based on graph neural networks except our recent work on the \ac{GAT} channel estimator, yet. The motivation behind using \acp{GAT} for channel estimation can be listed as (i) GATs can estimate all channel coefficients without an on-off switch mechanism, (ii) GATs' computational complexity is relatively low~\cite{velickovic_graph_2018}, (iii) GATs can be generalized over unobserved graphs due to the attention mechanism~\cite{lee_self-attention_2019}. Briefly, the \ac{GAT} channel estimator can reduce lower pilot overhead as it requires a single pilot signaling subframe to estimate channel coefficients regarding all \ac{RIS} elements. Furthermore, \ac{GAT} estimator can continue to perform well under variable channel conditions, thanks to the attention mechanism~\cite{tekbiyik2020channel}.}

Moreover, this study investigates \ac{BER} performance of the \ac{DtS} systems under more practical RIS designs unlike the prior works~\cite{basar2021reconfigurable, yildirim2020modeling, tekbiyik2020reconfigurable_sat} on wireless communications assisted by hypothetical RISs with continuous phase shift capability. While hypothetical RISs show almost independent performance from the channel model, the performance of practical RISs is significantly dependent on both the characteristics of the channel they operate and the performance of the channel estimation method, depending on their design. Therefore, the channel model and the channel estimation algorithm should be considered together when investigating the performance of practical RISs.

It has been shown in our previous work~\cite{tekbiyik2021energy} that it is possible to increase energy-efficiency by using \acp{RIS}. However, as \acp{RIS} strictly need \ac{CSI} to improve the received \ac{SNR}, we propose a channel estimation architecture based on \acp{GAT} in~\cite{tekbiyik2020channel} which overperforms \ac{LS} estimation. \textcolor{black}{Although~\cite{tekbiyik2020channel} made significant contributions as a proof-of-concept study, it was lacking in many aspects. This study is significantly extending the work in this field by presenting the following main contributions:}
\begin{enumerate}[{C}1]
\item \textcolor{black}{By comparing the channel estimation performance of the proposed \ac{GAT} model and the conventional \ac{DL} methods, it is shown that the \ac{GAT} shows a higher performance at low SNR. Moreover, the \ac{GAT} can \textcolor{black}{maintain} its performance much more than the conventional \ac{DL} methods under unobserved channel conditions. It should be noted that the proposed \ac{GAT} is less \textcolor{black}{time-consuming} for training due to its short epoch time compared to the conventional \ac{DL} methods.}
\item In addition to channel estimation, we design \ac{TDD} framework for uplink and downlink signaling of \ac{RIS}-assisted \ac{DtS} \ac{IoT} in order to use \textcolor{black}{partial channel reciprocity}. Due to the motion of satellites, the coherence time is short. Thus, the pilot signaling subframe must be extremely short duration while the channel estimation method should show high performance with the small number of pilot symbols. In this study, the proposed \ac{GAT} estimator uses only $16$ symbols for the pilot subframe.
\item The performance of \ac{RIS}-assisted \ac{DtS} \ac{IoT} is investigated for non-ideal \acp{RIS} with discrete and distinct phase sets as well as ideal \acp{RIS} when \ac{CSI} is acquired by the proposed \ac{GAT} channel estimator. \textcolor{black}{By considering the numerical results, the relation between \ac{RIS} design and the channel models is discussed.}
\end{enumerate}

The rest of this paper is organized as follows. \SEC{sec:preliminaries} introduces the basic background for \acp{GAT} and \ac{RIS}-assisted satellite communications. In \SEC{sec:system_model}, the system model is detailed for \ac{RIS}-assisted \ac{DtS} \ac{IoT} supported by \ac{GAT} channel estimator. \SEC{sec:ch_est} describes \textcolor{black}{the} channel estimation procedure from dataset generation to training by giving the related parameters for \ac{GAT}. In \SEC{sec:results}, numerical results under various \ac{RIS} configurations are discussed. Finally, \SEC{sec:conclusion} dwells on the open issues and concludes the study.

\section{Preliminaries}\label{sec:preliminaries}
In this section, the fundamentals for each part of the proposed \ac{RIS}-assisted satellite \ac{IoT} communication with \acp{GAT} channel estimator are introduced. First, we present the \ac{GAT} in detail. Then, \ac{RIS}-assisted satellite links are discussed. It should be noted that the notation used in this section is given for downlink; however, it can be used for uplink transmission without loss of generality.

\subsection{Graph Attention Networks}\label{sec:gat}

Graph neural networks have been recently proposed as a state-of-the-art solution for data that does not exhibit a grid-like structure while most deep learning methods are utilized for the data in the regular domain. \ac{GAT}, which is one of the graph neural networks, is prominent for inductive learning thanks to its attention mechanism. Inductive learning provides the generalization of a trained network over unobserved graphs. Considering the random nature of the propagation medium, \ac{GAT} is suitable to be utilized over unobserved channel states.

A \ac{GAT} consists of \acp{GAL} with $P$ input nodes denoted by $\mathbf{\vartheta}=\left\{\vec{\vartheta}_{1}, \vec{\vartheta}_{2}, \ldots, \vec{\vartheta}_{P}\right\}, \, \vec{\vartheta}_{i} \in \mathbb{R}^{F}$. $F$ stands for the number of features in each node. The output set of node features for \ac{GAL} can be shown by a new set $\mathbf{\vartheta}^{\prime}=\left\{\vec{\vartheta}_{1}^{\prime}, \vec{\vartheta}_{2}^{\prime}, \ldots, \vec{\vartheta}_{P}^{\prime}\right\}, \, \vec{\vartheta}_{i}^{\prime} \in \mathbb{R}^{F^{\prime}}$. Since the cardinality for the output and input might be different, the number of features is represented by $F^{\prime}$. The input properties of each node are transformed to higher-level properties by utilizing a linear transformation described by the weight matrix, $\mathbf{W} \in \mathbb{R}^{F\times F^{\prime}}$. Then, the attention mechanism, $a: \mathbb{R}^{F^{\prime}} \times \mathbb{R}^{F^{\prime}} \rightarrow \mathbb{R}$, is employed to evoke the self-attention on nodes. The attention coefficients are computed as follows:
\begin{align}
c_{i j}=a\left(\mathbf{W} \vec{\vartheta}_{i}, \mathbf{W} \vec{\vartheta}_{j}\right),
\end{align}
where $c_{i j}$ denotes the neighborhood between the $i$-th and $j$-th nodes in the graph. The attention coefficients reveal how much the features of the $j$-th node have an impact on the $i$-th node. By using a softmax function, the attention coefficients are normalized as given~\cite{bahdanau_neural_2016}
\begin{align}
\alpha_{i j}=\operatorname{softmax}_{j}\left(c_{i j}\right)=\frac{\exp \left(c_{i j}\right)}{\sum_{k \in \mathcal{N}_{i}} \exp \left(c_{i k}\right)},
\end{align}
where the neighborhood for $i$-th node is denoted by $\mathcal{N}_{i}$. The attention mechanism determines the normalized coefficients, $\alpha_{i j}$, as~\cite{velickovic_graph_2018} 
\begin{equation}
\alpha_{i j}=\frac{\exp \left(\operatorname{ReLU}\left(\mathbf{a}^{\top}\left[(\mathbf{X} \mathbf{W})_{i} \|(\mathbf{X} \mathbf{W})_{j}\right]\right)\right)}{\sum_{k \in \mathcal{N}(i)} \exp \left(\operatorname{ReLU}\left(\mathbf{a}^{\top}\left[(\mathbf{X} \mathbf{W})_{i} \|(\mathbf{X} \mathbf{W})_{k}\right]\right)\right)},
\end{equation}
where $\mathbf{X} \in \mathbb{R}^{P \times F}$ and $\mathbf{a} \in \mathbb{R}^{2 F^{\prime}}$ are node features and attention kernel, respectively. The convolution operation is performed over the graph network as follows\textcolor{black}{:}
\begin{equation}
\mathbf{Z}=\alpha \mathbf{X} \mathbf{W}+\mathbf{b},
\end{equation}
where $\mathbf{b}$ refers to the trainable bias vector. The inputs of this layer are the node attributes matrix $\mathbf{X} \in \mathbb{R}^{P \times F}$, the edge attributes matrix $\mathbf{E} \in \mathbb{R}^{P \times P \times S}$, and the binary adjacency matrix $\mathbf{A} \in\{0,1\}^{P \times P}$. Moreover, a pooling layer is employed to generalize graph convolution networks~\cite{lee_self-attention_2019}. Besides generalization, the pooling layer enables to decrease \textcolor{black}{in} the number of representations. As a result, it can be said that the pooling layer avoids the graph neural network to overfit. We employ only \textcolor{black}{a} global attention pooling layer in the proposed graph neural network. The output of global attention pooling for the input, $\mathbf{X}$, can be given as
\begin{align}
\mathbf{X}^{\prime}=\sum_{i=1}^{P}\left(\sigma\left(\mathbf{X} \mathbf{W}_{1}+\mathbf{b}_{1}\right) \odot\left(\mathbf{X} \mathbf{W}_{2}+\mathbf{b}_{2}\right)\right)_{i}.
\label{eq:x_prime}
\end{align}
In~(\ref{eq:x_prime}), $\sigma$ denotes the sigmoid function and $\odot$ is the broadcast elementwise product.

\subsection{RIS-assisted LEO Satellite Communications}\label{sec:ris_leo_system_model}

\begin{figure}[!t]
    \centering
    \includegraphics[width=\linewidth, page = 6]{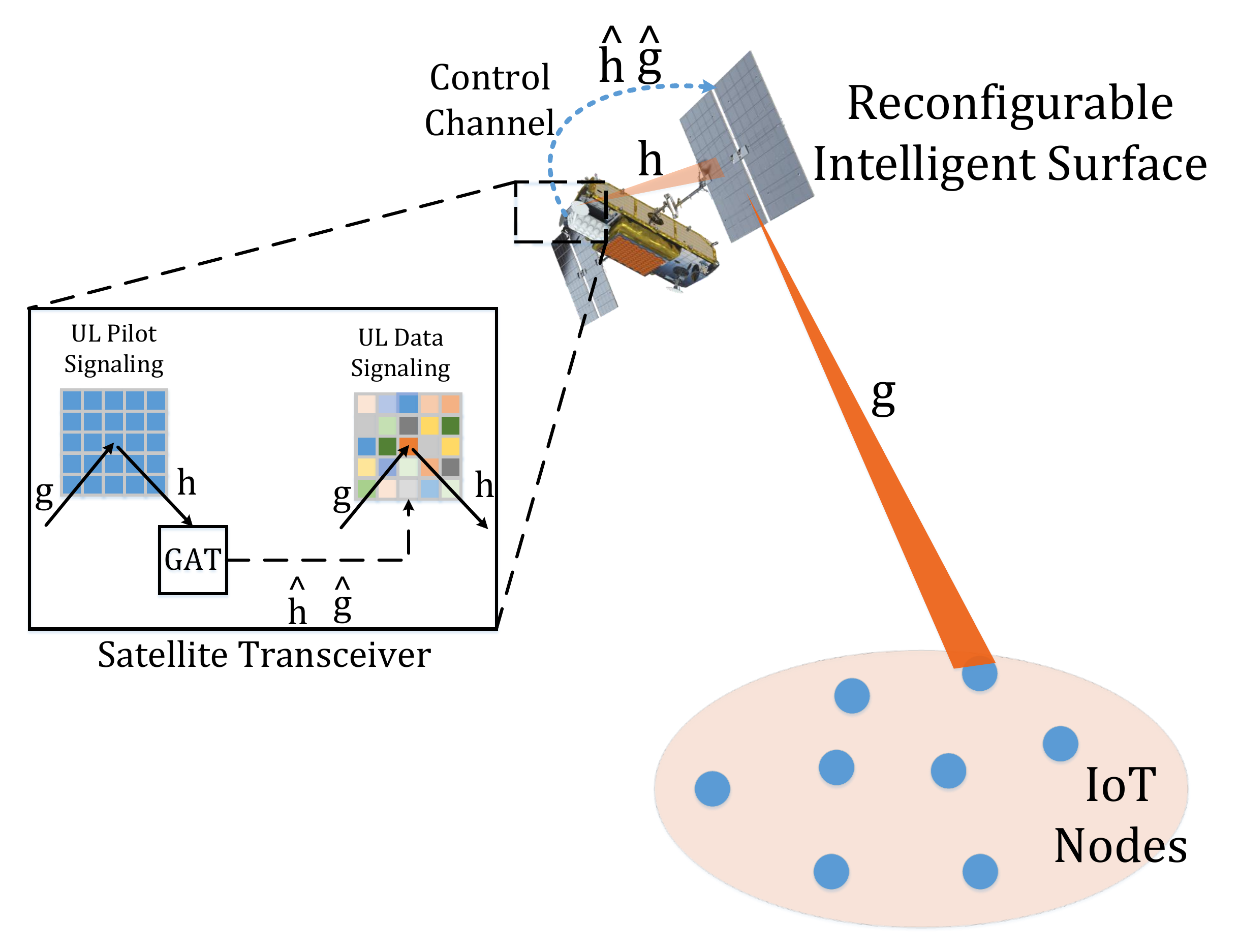}
    \caption{Direct-to-Satellite IoT communications assisted by \ac{RIS} with the \ac{GAT} channel estimator. The estimated channel state information is employed to reconfigure RIS elements.}
    \label{fig:ris_satellite}
\end{figure}

In this section, we introduce a system model of \ac{RIS}-assisted \ac{DtS} communications for \ac{IoT} networks. As known that relays can show higher performance when they become closer to the receiver or the transmitter, \ac{RIS} is deployed near the satellite antenna as illustrated in \FGR{fig:ris_satellite}. Thus, the system scheme provides a two-fold gain which is maximizing improvement and avoiding extra wireless \textcolor{black}{fronthaul} between transmitter and \ac{RIS} for \ac{CSI}. It is worth noting that it is assumed that the satellite antenna is aligned with the normal line of the \ac{RIS} to maximize the normalized radiation pattern~\cite{tekbiyik2021energy} \textcolor{black}{and the transmit antenna is in the far-field of the RIS to enable beamforming~\cite{tang2020wireless}}. Under this system model, the received signal reflected by \ac{RIS}, $r$, can be given as 
\begin{equation}
r =\sqrt{P_{t}\xi} \mathbf{g^\mathrm{T}} \mathbf{\Phi} \mathbf{h} x + w,
\end{equation}
where the transmitted signal with power $P_{t}$ and \ac{AWGN} at receiver are shown by $x$ and $w \thicksim \mathcal{C} \mathcal{N}\left(0, N_{0}\right)$, respectively. $\xi$ is the total path loss in \ac{RIS}-assisted communications, as detailed in~\cite{tekbiyik2021energy}. As the transmit antenna is near the \ac{RIS}, it is worthwhile to employ the near-field beamforming scheme against the attenuation due to atmosphere~\cite{tekbiyik2021energy}. $\mathbf{h}=\left[h_{1}, h_{2}, \ldots, h_{N}\right]$ and $\mathbf{g}=\left[g_{1}, g_{2}, \ldots, g_{N}\right]$ stand for the channel coefficient vectors, where $h_i = \beta_i \mathrm{e}^{j \theta_{i}}$ and $g_i = \rho_i \mathrm{e}^{j \nu_{i}}$. In this study, the amplitude coefficients, $\beta_i$ and $\rho_i$, are assumed to follow the Rician distribution with the shape parameter of $K = 10$ to evaluate a slight multipath fading~\textcolor{black}{\cite{letzepis2008capacity, you2020massive, jung2022performance}}. $\theta_i$ and $\nu_i$ denote the phase response of the channels regarding the $i$-th \ac{RIS} element. Also, $\mathbf{\Phi}$ is the \ac{RIS} phase shift matrix given as
\begin{equation}
\Phi=\operatorname{diag}\left\{A_{1} \mathrm{e}^{-j \phi_{1}}, \ldots, A_{N} \mathrm{e}^{-j \phi_{N}}\right\},
\end{equation}
where $\phi_{i}$ and $A_{i}$ \textcolor{black}{denote} the phase and amplitude response of $i$-th \ac{RIS} element\textcolor{black}{, respectively}. It is worth mentioning that \ac{RIS} is assumed as a lossless device throughout this study; hence, $A_{i} = A = 1, \; \forall i$. Besides, $\phi_{i}$ is determined through the estimated channel coefficients, $\hat{h}_i$ and $\hat{g}_i$, as follows\textcolor{black}{:}
\begin{align}
    \phi_{i} = \hat{\theta}_i + \hat{\nu}_i,
\end{align}
where $\hat{\theta}_i$ and $\hat{\nu}_i$ are the phase shift values estimated by the \ac{GAT} channel estimator.

Then, the instantaneous effective \ac{SNR}, $\gamma^{\mathrm{eff}}$, is given as follows\textcolor{black}{:}

\begin{equation}
\gamma^{\mathrm{eff}}=\frac{P_{t}\xi\left|\left(\sum_{i=1}^{N} \beta_{i} \rho_{i} e^{j \psi_i} \right)\right|^{2}}{N_{0}},
\end{equation}
where $\psi_i = \phi_i - \theta_i - \nu_i$ and $\psi_i = 0$ for the ideal channel estimation.

\section{Direct-To-Satellite IoT Communications}\label{sec:system_model}

\begin{figure}[!t]
    \centering
    \includegraphics[width=\linewidth, page=3]{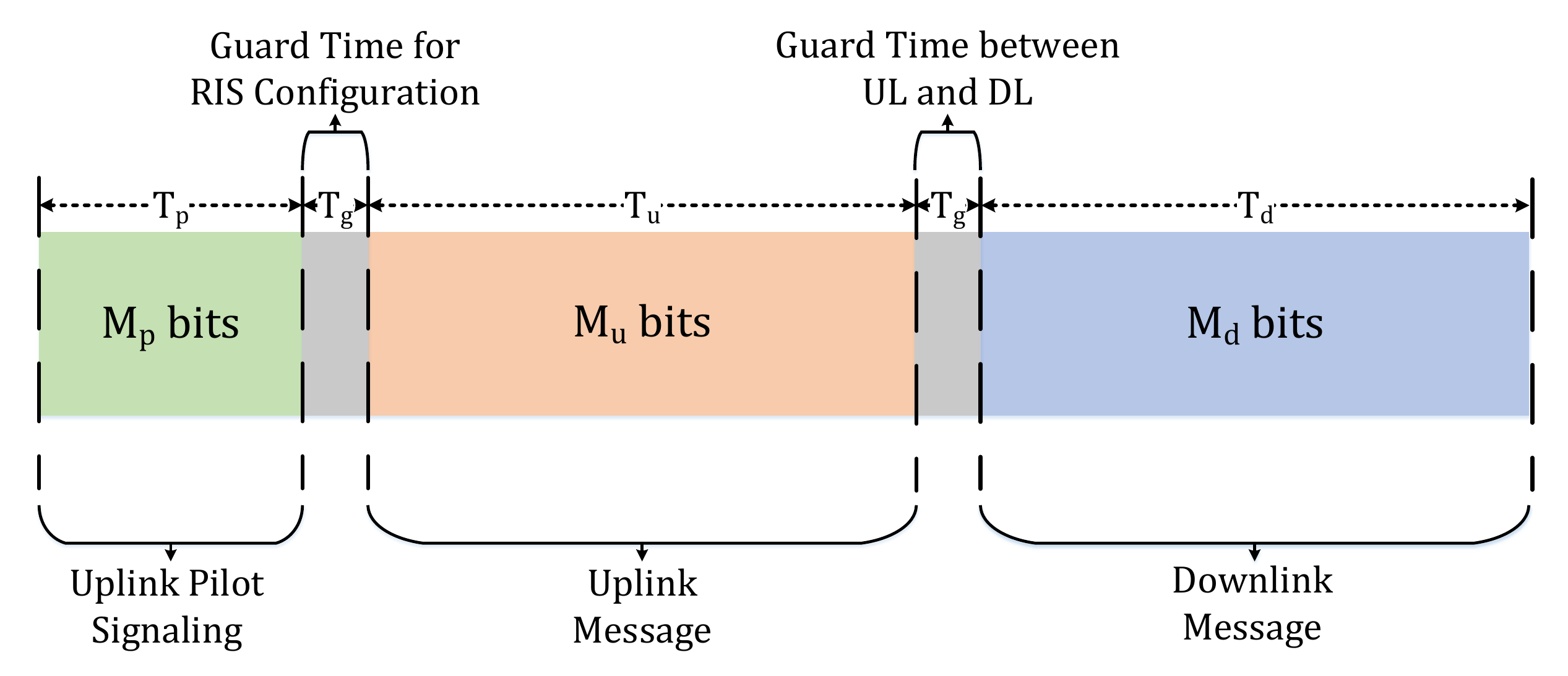}
    \caption{Each \ac{TDD} frame consists of uplink pilot signaling, uplink message and downlink message subframes. The frame starts with pilot signaling to estimate channel coefficients regarding to \ac{RIS} elements.}
    \label{fig:signaling_scheme}
\end{figure}

In this section, \ac{LEO} satellite-enabled \ac{IoT} communications system model is introduced. Before detailing the system model, we would describe the motivations behind the proposed system model. 
In this study, \textcolor{black}{a narrowband modulation-based} physical layer is adopted. This scheme utilizes a signal with a carrier which has very narrow bandwidth. By employing this scheme, it is possible to design low-complex transceivers~\cite{anteur2015ultra}. Thus, the cost for \textcolor{black}{the} transceiver part of \ac{IoT} devices can be reduced. Moreover, it should be noted that narrow band modulation can show resistance to noise and interference due to its high power spectral density~\cite{lassen2014long}. Therefore, it is possible to employ ultra-narrow band signals on shared frequency bands. Another appealing feature of ultra-narrow band modulation is that it enables long-range communication link with low-power consumption~\cite{xiong2015low}. As a result, ultra-narrow band modulation techniques can provide to employ low-complex transceiver design in both satellite and \ac{IoT} devices. Moreover, it can increase battery life by using low power for transmission and reception. However, besides all the appealing features of ultra-narrow band, data rates supported by the ultra-narrow band signals are very low. 

Two methods can be proposed to increase the data rate: increasing the modulation degree and/or increasing the bandwidth. Since the required received \ac{SNR} value raises with increasing the modulation degree, the transmission power should also be \textcolor{black}{enhanced}. This reduces energy-efficiency. Since increasing the bandwidth increases the in-band noise in the receiver, it is evident to increase the transmitted power. More importantly, the equalizer is required in the receiver in order to mitigate multipath fading. To improve energy-efficiency, we have recently proposed RIS-assisted satellite \ac{IoT} communications in~\cite{tekbiyik2021energy}. In this study, we revise the link budget analysis and \textcolor{black}{achievable} capacity for RIS-assisted satellite \ac{IoT} communication under the assumption of the perfect channel state information. \textcolor{black}{\cite{tekbiyik2021energy}~shows that \acp{RIS} can achieve significant improvement in energy-efficiency for both uplink and downlink. Due to improved energy-efficiency, the transmit power is relatively low in both links. Furthermore, the hardware complexity of IoT devices might decrease because of no shortage of complex amplifiers. \textcolor{black}{Although it might be thought that there is no need for an equalizer in narrowband communication, equalization is required for finding channel gain to compensate \textcolor{black}{for} channel effect, and then demodulate the received signals. Several studies such as~\cite{pancaldi2008single, benvenuto2009single, zhang2017channel, popli2018survey, ali2019frequency} have already addressed the channel equalization for narrow band communication schemes. However, the computational complexity of the proposed methods is high and they either need high computational capacity or consume high power. Noting that \acp{RIS} are devices that can process signals at rather than transmitter and receiver, the transceiver complexity can be reduced. Moreover, \acp{RIS} can perform equalizer tasks over a communication medium instead of transceivers~\cite{basar2019wireless, arslan2021over}.}} 

We adopt \ac{TDD} for uplink and downlink communications to exploit \textcolor{black}{partial channel reciprocity}. By uplink pilot signaling, the \ac{CSI} is estimated in the satellite by utilizing \acp{GAT} as detailed \textcolor{black}{in~\SEC{sec:ch_est}}. Each \ac{TDD} frame includes $M_p$ uplink pilot symbols, $M_u$ uplink message symbols, and $M_d$ downlink message symbols. Furthermore, \textcolor{black}{the} \ac{TDD} frame consists of guard intervals for \ac{RIS} configuration and \textcolor{black}{avoids} interference between uplink and downlink signals. \textcolor{black}{The} \ac{TDD} frame is illustrated in~\FGR{fig:signaling_scheme}. It is worth noting that the total length of \ac{TDD} subframe is less than the coherence interval. The $M_p$-length pilot symbols are selected as a \ac{PN} sequence generated by the polynomial given by $x^4+x^2+1$. Using \ac{PN} sequence provides time synchronization between satellite and \ac{IoT} devices and the detection of the starting point of \ac{TDD} frame. Additionally, \ac{PN} sequence might be utilized to identify \ac{IoT} device by assigning unique sequences for each device.

\begin{figure*}[!t]
    \centering
    \includegraphics[width=0.8\linewidth, page=2]{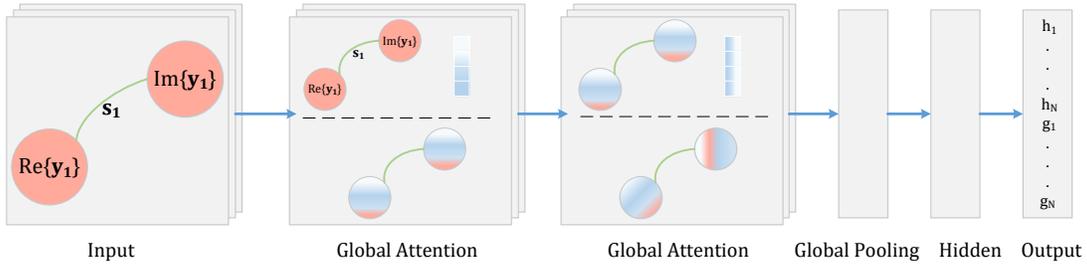}
    \caption{The illustration for the proposed \acl{GAT} including two consecutive \acl{GAT}s and global attention pooling. The real and imaginary parts of \textcolor{black}{the} received signal, $\mathbf{y}$, are assigned to attributes of two nodes. The edge attributes are set as the pilot symbols, $\mathbf{s}$.}
    \label{fig:graph}
\end{figure*}

After channel estimation, \ac{RIS} elements are configured to maximize the received \ac{SNR}. Then, the uplink signal can be demodulated without a complex equalizer because \ac{RIS} can mitigate the random behavior of wireless channels. For downlink communication, the configured \ac{RIS} is illuminated by the transmitter antenna on the satellite. The reflected signal reaches the \ac{IoT} device with a high \ac{SNR}. In consequence, \ac{IoT} device can demodulate the reflected signal with low power consumption~\cite{tekbiyik2021energy}.

This scheme mandates \textcolor{black}{having} an uplink pilot signal for channel estimation before downlink, and therefore sending a message in the downlink depends on the presence of uplink communication. However, when a satellite needs to send a downlink message without waiting for a message from the uplink, the target \ac{IoT} devices can be evoked by operating \ac{RIS} in broadcasting mode. Evoked \ac{IoT} sends uplink pilot sequence for channel estimation. 

It is important to note that the proposed scheme is well suited to random time multiple access techniques such as ALOHA and time-slotted ALOHA or \ac{TDMA}. However, random frequency multiple access can be supported by providing extra functionalities to the system. For example, a wideband \ac{RIS} design is required for multiple access based on frequency domain.

\section{Channel Estimation Procedure}\label{sec:ch_est}

\begin{table}[]
\centering
\caption{Summary of the dataset parameters employed during training and test.}
\begin{tabular}{lcc}
\toprule \toprule
\multicolumn{1}{l}{\textbf{Parameters}} & \multicolumn{1}{c}{\textbf{Training}} & \multicolumn{1}{c}{\textbf{Test}} \\  \midrule
\multicolumn{1}{l}{PN Polynomial}       & $x^4+x^2+1$                           & $x^4+x^2+1$                       \\
\multicolumn{1}{l}{Modulation}       & BPSK                           & BPSK                       \\
\multicolumn{1}{l}{\# of Samples per SNR}       & 1000                                  & 500                               \\
\multicolumn{1}{l}{SNR (dB)}            & -30:2:0                               & -30:2:10                          \\
\multicolumn{1}{l}{K}                   & 10                                    & 10                   \\
\multicolumn{1}{l}{$M_p$}                   & 16                                    & 16                   \\
\multicolumn{1}{l}{$N$}                   & 16, 32, 64                                   & 16, 32, 64                   \\ \bottomrule \bottomrule
\end{tabular}
\label{tab:parameters}
\end{table}

This section is devoted to introducing the proposed channel estimation methodology with \acp{GAT}. As mentioned earlier, the attractive functionality of \acp{GAT} that allows it to generalize to graphs that have not been completely observed during training~\cite{velickovic_graph_2018} makes it a healer for channel estimation in \ac{RIS}-assisted communications. Unlike the channel estimation methods such as~\cite{mishra2019channel, elbir2020deep} based on switching \ac{RIS} elements, the proposed \ac{GAT} channel estimator is able to acquire all channel coefficients regards to RIS elements in a single pilot signaling subframe. Due to \textcolor{black}{the} motion of \ac{LEO} satellites, channel coefficients rapidly change. Therefore, accurate channel estimation in \ac{LEO} satellites requires a generalizable network to keep estimation performance stable. As mentioned above, it is possible to generalize \acp{GAT} to unobserved graphs~\cite{velickovic_graph_2018}.

\subsection{Dataset Generation}\label{sec:dataset_generation}

The channel coefficients, $\mathbf{h}$ and $\mathbf{g}$, are estimated at the satellite by using uplink pilot signaling. To do this, $M_p$-length pilot subframes are created by using \ac{PN} sequence \textcolor{black}{generated} by the polynomial $x^4+x^2+1$. Assigning different \ac{PN} sequences to each \ac{IoT} device allows both identifying the device at the satellite side and \textcolor{black}{maintaining} synchronization. \textcolor{black}{Since the DtS IoT systems rely on narrowband modulation schemes~\cite{fraire2019direct} and the recent studies~\cite{hofmann2019ultranarrowband, hofmann2019direct, kim2019performance} proposed \ac{BPSK}-based waveforms for both GEO and LEO IoT services, we utilize \ac{BPSK} for both pilot and message \textcolor{black}{signaling}.}

All meta-atoms are switched on with zero-phase shift, scilicet unitary phase shift matrix. To include a slight multipath effect in the dataset, we select $\beta_i$ and $\rho_i$ as Rice distributed with the shape parameter $K=10$. It is worth noting that the \ac{GAT} channel estimator can \textcolor{black}{maintain} the estimation performance under different multipath characteristics as shown in~\cite{tekbiyik2020channel}. By setting the parameters given above, the dataset has been created. The dataset consists of the input regarding \textcolor{black}{the} received signal, $\mathbf{X}$, and the adjacency matrix, $\mathbf{A}$, for the graph network including the real and imaginary parts of \textcolor{black}{the} channel coefficients. $\mathbf{X}$ and $\mathbf{A}$ are expressed as follows\textcolor{black}{:}
\begin{align}
    \mathbf{X} &= \left[\mathrm{Re}\{\mathbf{y}\}; \, \mathrm{Im}\{\mathbf{y}\}\right] \\
    \mathbf{A} &= \begin{bmatrix}
    0 & 1 \\
    1 & 0
\end{bmatrix}.
\end{align}
$\mathbf{A}$ denotes that a single edge connects two nodes as depicted in \FGR{fig:graph}. \textcolor{black}{It is worth noting that this crucial and unconventional approach enable\textcolor{black}{s} to track the phase changes during the pilot \textcolor{black}{signaling} and reveal the channel characteristics.} Also, the dataset includes the weight matrix of the edge for the $j$-th \textcolor{black}{non-zero} element of \textcolor{black}{the} adjacency matrix given as
\begin{align}
\mathbf{E}_j = \mathbf{s}, \, \mathbf{E} \in \mathbb{C}^{2\times2\times M}.
\end{align}
The label vector, $\mathbf{y}$, including the known channel coefficients is generated as 
\begin{align}
\mathbf{y} = \left[h_1, h_2, \cdots, h_N, g_1, g_2, \cdots, g_N\right]^{\mathrm{T}}.
\end{align}

The training dataset consists of $1000$ input samples for each \ac{SNR} level within -$30$ dB and $0$ dB. The step size for \ac{SNR} levels is $2$ dB. The total number of input samples in the training dataset is $16000$ for each $N$, and \ac{SNR} values. The training dataset has been divided into two parts: training and validation with the rate of $4:1$. \TAB{tab:parameters} summarizes the parameters that are used during the dataset generation.

\begin{table}[!t]
\centering
\caption{The parameters and layout for the proposed \ac{GAT} channel estimator.}
\begin{tabular}{ccc}
\toprule \toprule
\multicolumn{2}{c}{\textbf{Layers}}   & \textbf{Dimensions}  \\ \midrule
\multirow{3}{*}{Inputs}      & $\mathbf{X}$      & $2\times M_p$          \\ \cmidrule{2-3}
                             & $\mathbf{A}$      & $2\times 2$          \\ \cmidrule{2-3}
                             & $\mathbf{E}$      & $2\times 2 \times M_p$ \\ \midrule
\multirow{1}{*}{Labels}      & $\mathbf{y}$      & $4N \times 1$   \\ \midrule             
\multicolumn{2}{c}{Graph Attention 1} & $2\times 128$        \\
\multicolumn{2}{c}{Graph Attention 2} & $2\times 32$         \\
\multicolumn{2}{c}{Global Attention Pool}   & 128                  \\
\multicolumn{2}{c}{Dense}             & $4N$    \\ \toprule

\multicolumn{2}{c}{\textbf{Parameters}}   & \textbf{Values}  \\ \midrule
\multicolumn{2}{c}{Activation} & ReLU  \\
\multicolumn{2}{c}{Optimizer} & Adam  \\
\multicolumn{2}{c}{Loss} & MSE  \\
\multicolumn{2}{c}{Learning Rate} & 1e-3  \\
\multicolumn{2}{c}{L$_2$ Regularization} & 5e-4  \\
   
\bottomrule   \bottomrule 
\end{tabular}
\label{tab:gat_layout}
\end{table}

\subsection{\ac{GAT} Model and Training}

This section details the parameters of the proposed \ac{GAT} model that is implemented by using Spektral~\cite{grattarola2020graph}. The model consists of two consecutive \acp{GAL}. The first and second layers have 128 and 32 output channels, respectively. Each layer employs the ReLU activation function. The size of the input becomes $P = 2$, $F = M$, and $S = M$. Following \acp{GAL}, a global attention pooling layer is utilized to avoid the model overfitting by decreasing the number of representations. Moreover, it is worth noting that each \ac{GAL} dropouts \textcolor{black}{fifty percent} of the representations in order to reduce the model complexity as well avoiding overfitting.

Besides dropouts, the network employs $\mathrm{L}_2$ regularization. The learning flow through the network is terminated by a hidden layer with $4N$ neurons. In accordance with the nature of the channel estimation problem, the loss function is chosen as \ac{MSE}. To minimize the loss function, ADAM optimizer with a learning rate of $10^{-3}$ is employed when compiling the network. Although the number of epochs is determined as $20$, the early stopping is activated to keep training time short if the loss function does not decrease for $5$ epochs. \TAB{tab:gat_layout} summarizes the \ac{GAT} parameters and inputs.

\section{Numerical Results and Discussions}\label{sec:results}

\begin{figure}[!t]
    \centering
    \includegraphics[width=\linewidth]{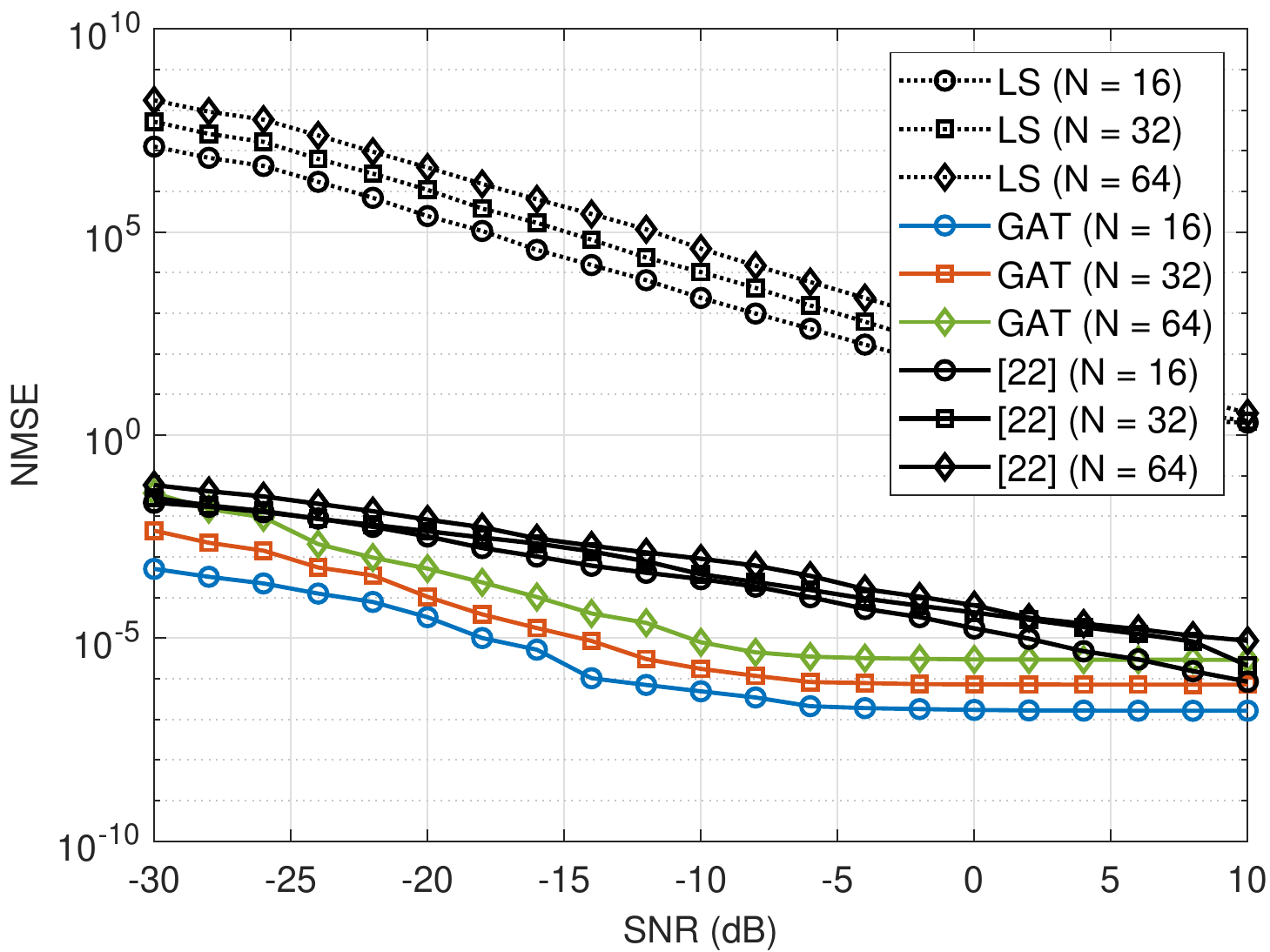}
    \caption{\textcolor{black}{\ac{NMSE} performance of the proposed \ac{GAT}-aided RIS-assisted satellite-to-IoT cascaded channel estimation versus the \acp{SNR}, the number of \ac{RIS} elements, $N$ for $M_p = 16$ and $K = 10$. It should be noted that $M_p = 256$ for~\cite{chen2019channel}.}}
    \label{fig:nmse}
\end{figure}

\begin{figure}[!t]
    \centering
    \includegraphics[width=\linewidth]{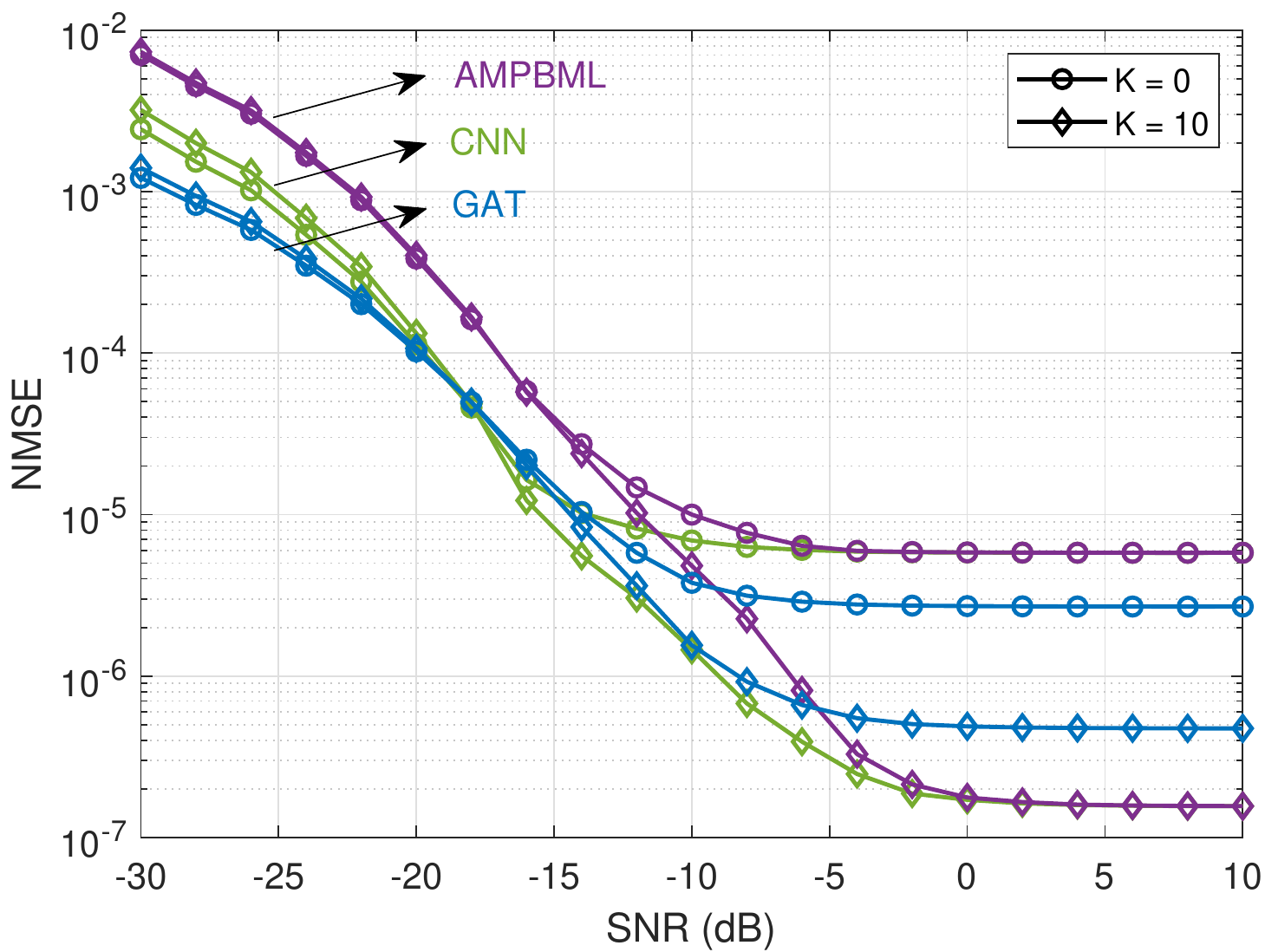}
    \caption{\ac{NMSE} performance of the proposed \ac{GAT} and other DL methods, the number of \ac{RIS} elements, $N = 16$ for $M_p = 16$ and $K = 0$ and $10$. It is observed that the proposed GAT shows a higher performance than the conventional DL methods under changing channel conditions. These results emphasize the importance of attention mechanisms and inductive learning.}
    \label{fig:nmse_dl_comparison}
\end{figure}

\begin{figure}[!t]
    \centering
    \includegraphics[width=\linewidth]{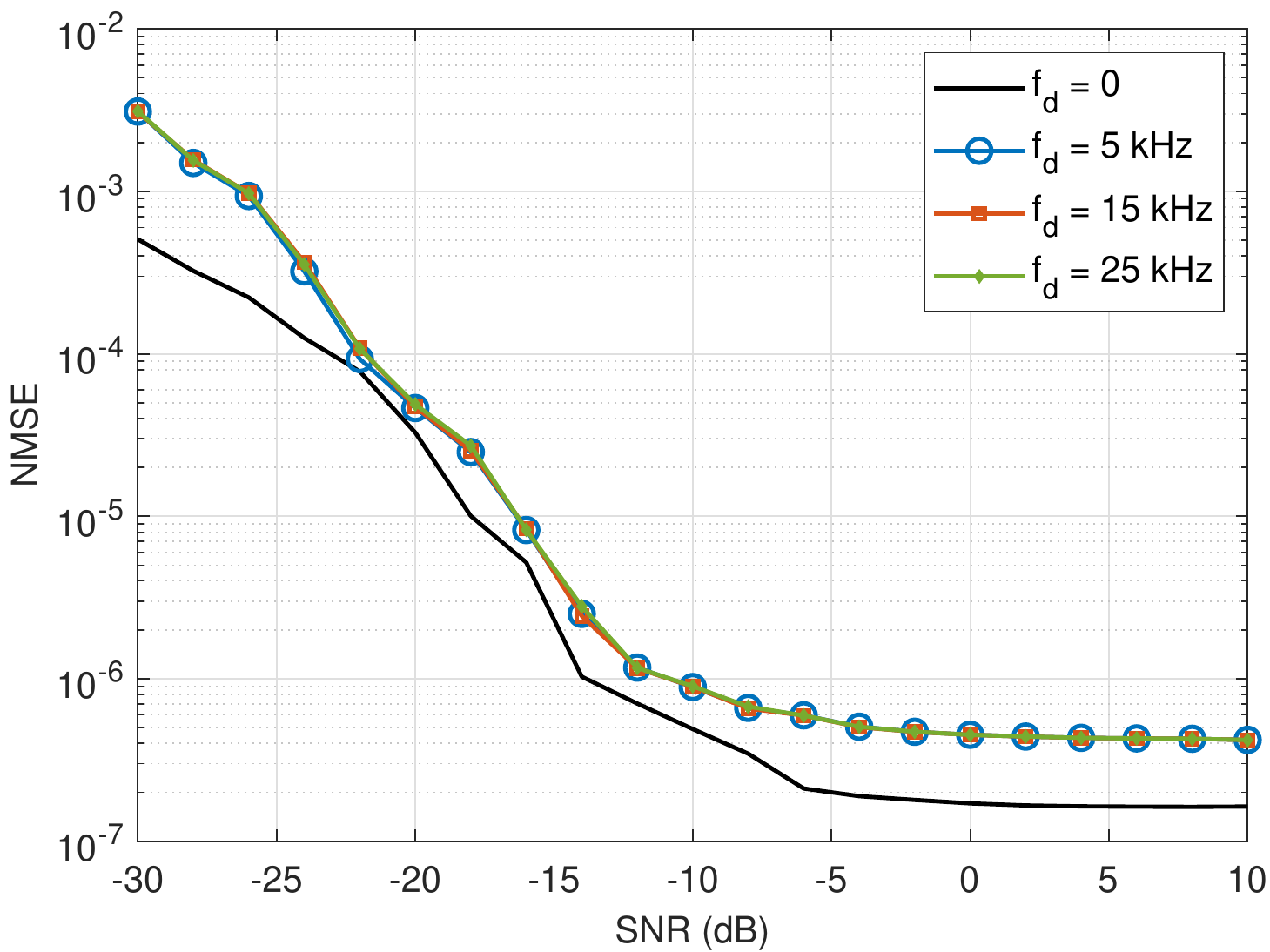}
    \caption{\textcolor{black}{\ac{NMSE} performance of the proposed \ac{GAT} under Doppler shift which observed in LEO satellite communication systems due to motion of satellites. The residual Doppler shift after compensation is denoted by $f_d$. It is observed that the proposed GAT-based channel estimator can almost maintain the performance under the Doppler shift although the training set does not consist of any Doppler effects. These results also denote the importance of attention mechanisms and inductive learning.}}
    \label{fig:gat_doppler}
\end{figure}

In this section, the performance of \ac{RIS}-assisted satellite \ac{IoT} communications is investigated under \ac{GAT} channel estimation procedure detailed in the previous section. As \cite{tekbiyik2020channel} denotes that the \ac{GAT} estimator is able to keep the same estimation performance for the decreasing number of pilot symbols. Therefore, we keep the length of pilot signaling as much as possible to avoid pilot contamination. In this study, the number of pilot symbols, $M_p$ is selected as $16$. Moreover, the \ac{GAT} estimator is robust to changes in fading statistics, due to attention mechanism~\cite{tekbiyik2020channel, velickovic_graph_2018}. In this study, we consider only Rician fading with the shape parameter of $K = 10$ to allow slight \ac{NLOS} components in the channel model \textcolor{black}{as given in~\cite{letzepis2008capacity, you2020massive, jung2022performance}}. The simulation parameters are summarized in the test column of \TAB{tab:parameters}. 

Firstly, the channel estimation performances for both $\mathbf{h}$ and $\mathbf{g}$ are considered in \FGR{fig:nmse}. The proposed \ac{GAT} estimator outperforms \ac{LS} estimation in terms of \ac{NMSE}. It is seen that both methods require an additional $3$ dB \ac{SNR} when the number of meta-atoms is doubled. But, it should be noted that the required transmit power reduces $3$ dB since doubling the number of elements decreases the required transmit power \textcolor{black}{by $6$ dB}~\cite{tekbiyik2020reconfigurable_sat, tekbiyik2020reconfigurable}. However, the proposed \ac{GAT} estimator is overperformed compared to \ac{LS} estimator. Furthermore, even though the training set does not include \ac{SNR} values between $0$ and $10$ dB, the \ac{NMSE} performance does not deteriorate as seen in \FGR{fig:nmse}. The \ac{NMSE} of \ac{GAT} converges to $10^{-7}$ for increasing \ac{SNR} value. \textcolor{black}{Moreover, \FGR{fig:nmse} denotes that the proposed method can show higher performance compared to \cite{chen2019channel} while the proposed method uses a one-sixteenth long pilot subframe of \cite{chen2019channel}.}

Besides \textcolor{black}{non-DL methods}, we \textcolor{black}{evaluate} conventional \ac{DL} methods for the same problem. In this \textcolor{black}{study}, we discuss the results of two of the recently proposed conventional \ac{DL} methods, which have the highest performance, compared with the \ac{GAT}. AMPBML has been proposed in~\cite{ma2020machine} for beam alignment in mmWave massive \ac{MIMO}. Another method based on \ac{CNN} has been proposed in~\cite{elbir2020deep} for channel estimation in \ac{RIS}-assisted mmWave communications. The performance of the models, which have been trained in the case of $K=10$, is investigated under more challenging conditions by also considering the Rayleigh channel (i.e., $K=0$) in the test dataset. In \FGR{fig:nmse_dl_comparison}, it is shown that the proposed GAT overperforms compared to the conventional methods at low SNR. Although GAT performs slightly lower than conventional methods at high SNR, it maintains its performance much better than the conventional DL methods in changing channel conditions. 
Owing to the GATs’ inductive learning ability over unobserved cases, GAT outperforms the aforementioned machine learning methods when the test data contains different channel conditions than the training dataset. These results reveal how important the attention mechanism is in the cases unobserved during training. Further, more detailed studies are needed on how the conventional methods perform under changing conditions due to a lack of inductive learning capability. As stated above, inductive learning due to the attention mechanism in \acp{GAT} can allow the network to generalize over unobserved graphs which can mean unobserved channel conditions in this study. It is worth noting that training complexity is another important factor besides performance in DL methods. From this point of view, it is an important advantage to train the proposed GAT model in a shorter time compared to other models. For example, AMPBML and CNN each require $1.8$ and $3$ times the epoch duration of GAT, respectively. In brief, the proposed GAT both shows a higher performance with increased robustness under changing conditions and has less computational complexity.

\textcolor{black}{Another crucial point in LEO-satellite communications systems that differs from terrestrial point-to-point communication systems is the Doppler effect because of the high mobility of satellites. Therefore, it should be carefully investigated in channel estimation performance since the Doppler shift changes through its orbit (i.e., elevation angle)~\cite{nguyen2021tcp}. From the point of channel estimation in LEO satellite systems, employing separate models for each Doppler shift or elevation angle is not practical and feasible for channel estimation; hence, the model must be attainable under changing conditions that have not been observed during the training process. We discuss and denote that the proposed method can maintain the performance under changing conditions such as small-scale fading and SNR above. Similar to what changes in small-scale fading affect the received signal phase, the Doppler shift can accumulate phase error on the received signal. Intuitively, it might be said that the proposed method can maintain the performance under the Doppler effect by considering the previous results. Besides the results discussed previously, we investigate the performance of the GAT-based estimator under the Doppler shift.} 

\textcolor{black}{As given in~\cite{nguyen2021tcp}, the maximum Doppler shift for LEO satellite at $600$ km altitude is ranging in between -$50$ and $50$ kHz. It should be noted that the higher the maximum Doppler shift is observed, the higher the altitude. Also, as the Doppler shift is monotonic and predictable, it is possible to compensate Doppler shift observed in LEO satellite communications~\cite{kodheli2020random}. For example, in~\cite{lin2016map}, maximum a posterior estimator was proposed to exploit the Doppler shift in LEO satellite communication systems. However, residual Doppler shift after compensation can affect the performance of the system and it should be investigated.}

\textcolor{black}{First, we generate a test dataset including signal samples that observed the Doppler effect with residual Doppler shifts, $f_d$, of $5$, $15$, and $25$ kHz after Doppler compensation~\cite{lin2016map}. Then, the model trained with data without any Doppler impact is tested with the new dataset. The results are shown in~\FGR{fig:gat_doppler} compared to the case without the Doppler shift (i.e., $f_d = 0$). As seen in~\FGR{fig:gat_doppler}, the impact of the Doppler shift on the channel estimation performance is very limited likewise the results in~\FGR{fig:nmse_dl_comparison}.  The magnitude of Doppler shifts has no remarkable impact on the performance. The results exhibit generalizability, which is the most desired feature for DL applications under changing conditions, of the proposed method.}

\begin{figure}[!t]
    \centering
    \includegraphics[width=\linewidth]{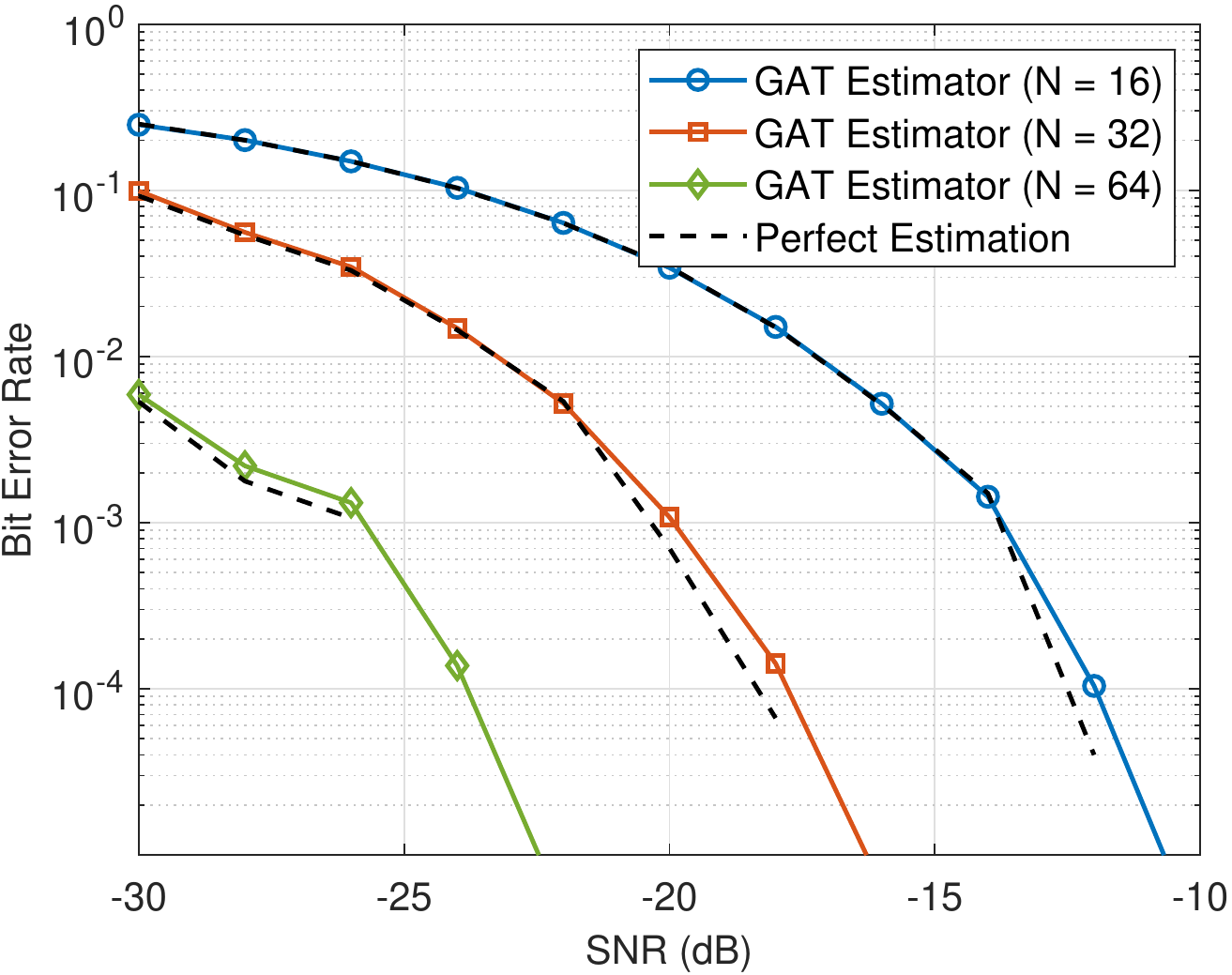}
    \caption{\ac{BER} performance of RIS-assisted satellite \ac{IoT} communications with \ac{BPSK} signaling under the ideal channel estimation and GAT-based estimation.}
    \label{fig:ber}
\end{figure}

Next, the error rate performance of the proposed \ac{RIS}-assisted \ac{DtS} \ac{IoT} system is investigated by considering the channel estimation error resulting from \ac{GAT} estimator. The channel estimation and then \ac{BER} analysis are performed for the number of \ac{RIS} elements of $16$, $32$, $64$. In this case, the \ac{RIS} is assumed as \textcolor{black}{a} continuous phase. The simulation results under both the perfect estimation and \ac{GAT} estimation are given in~\FGR{fig:ber}. It is observed that \ac{BER} performance in the case of \ac{GAT} estimator is almost same \textcolor{black}{as} the perfect estimation at the low \ac{SNR} region. As the \ac{SNR} value increases, there is a very slight degradation in the error performance due to the non-perfect estimation of the \ac{GAT}. Moreover, \FGR{fig:ber_upper_lower} denotes the confidence intervals for trained \ac{GAT} estimators. In each training phase, the model parameters and training data are kept the same. However, random initialization and randomness in the optimizer give rise to different trained models with distinct weight matrices. Intuitively, the best case is the almost same as the \ac{BER} results in \textcolor{black}{the} case of the perfect estimation. But, the increasing number of \ac{RIS} elements makes the confidence interval larger by resulting \textcolor{black}{in} the worst channel estimation error. 

\begin{figure}[!t]
    \centering
    \includegraphics[width=\linewidth]{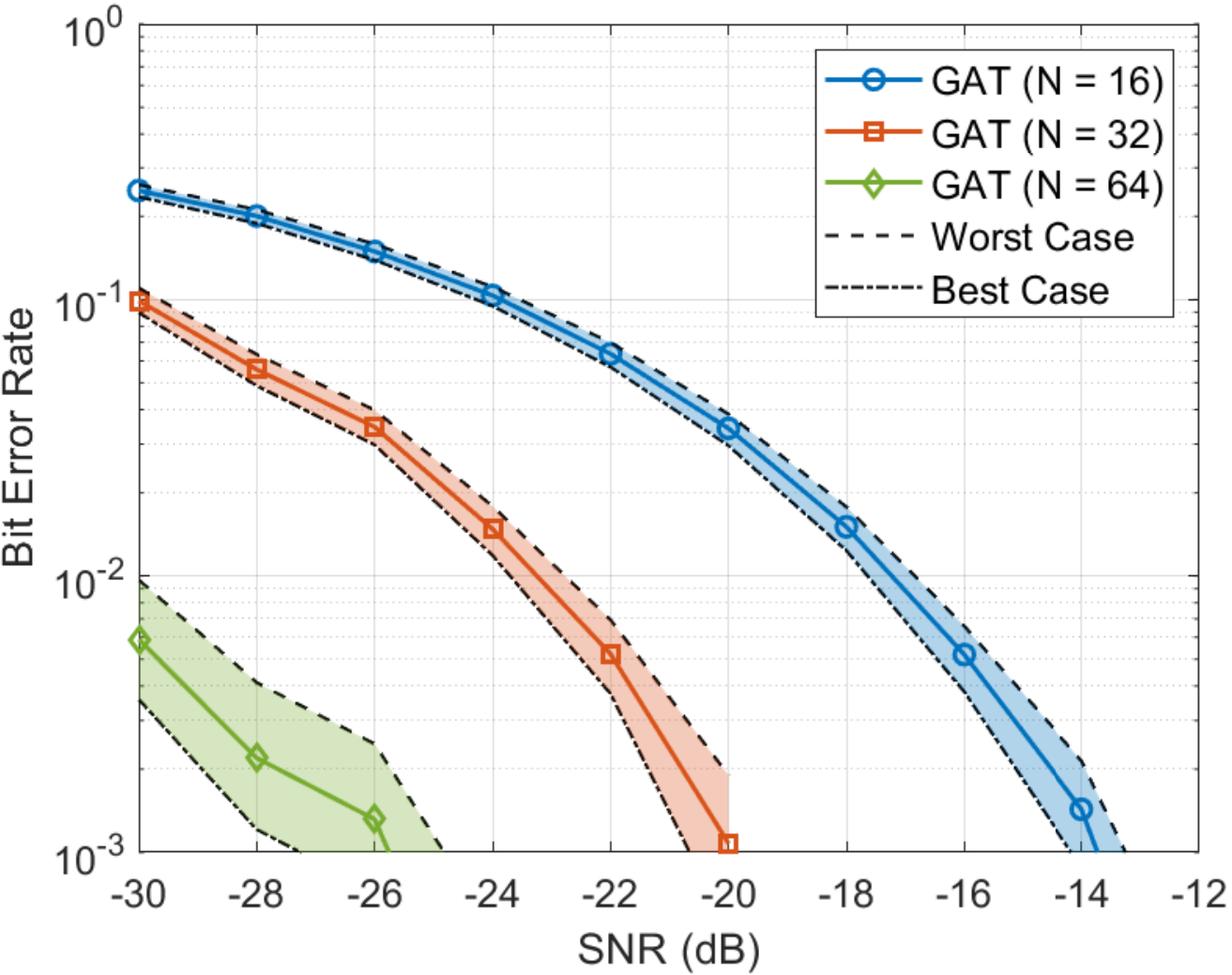}
    \caption{The confidence interval of \ac{BER} performance for training procedures. The GATs are trained without any change in the parameters and training sets.}
    \label{fig:ber_upper_lower}
\end{figure}

\begin{figure}[!t]
    \centering
    \includegraphics[width=\linewidth]{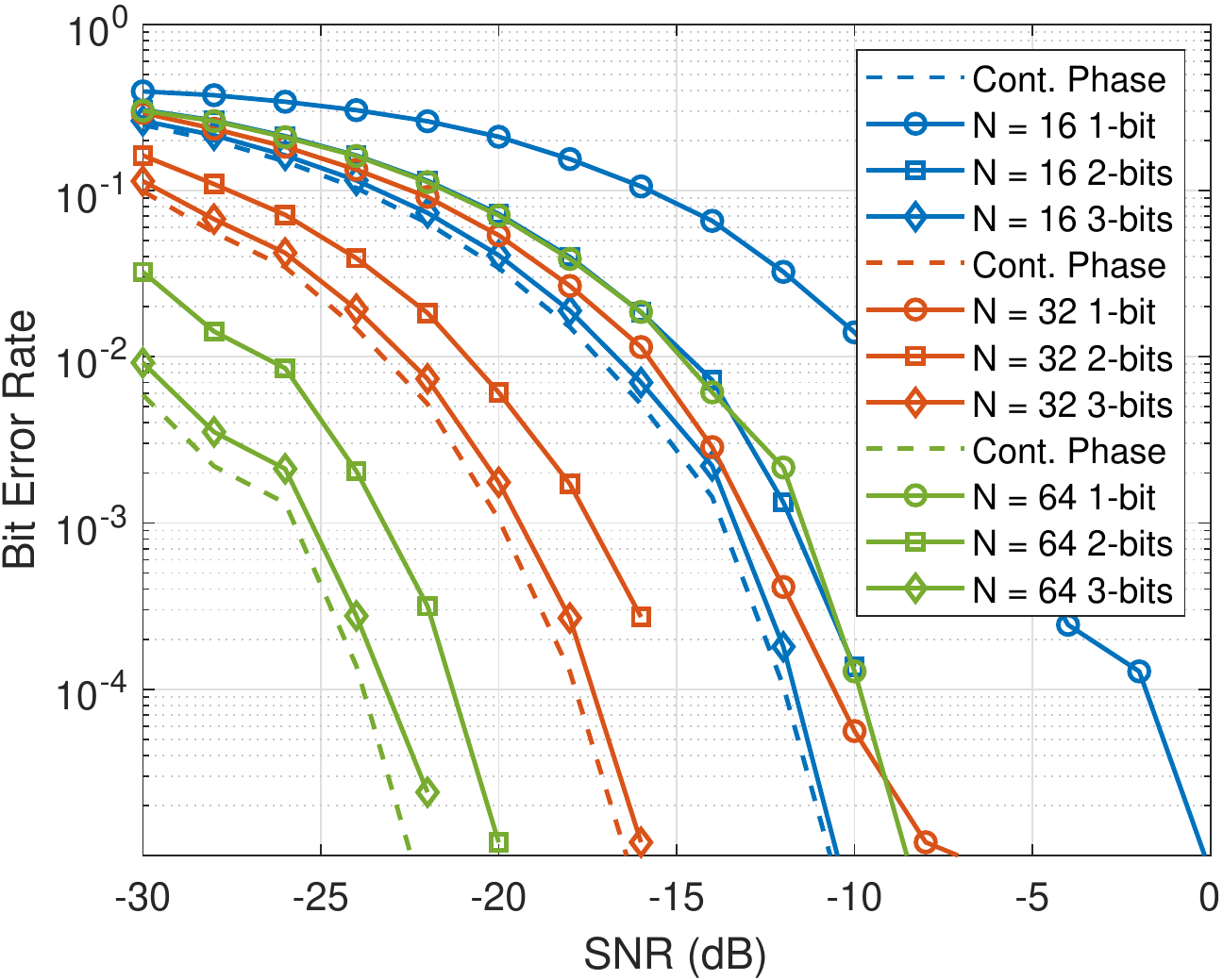}
    \caption{\ac{BER} performance of discrete phase RIS-assisted satellite \ac{IoT} communications with \ac{BPSK} signaling. Increasing the number of quantization levels proportionally improves the error rate performance of the system.}
    \label{fig:ber_discrete}
\end{figure}

Although \acp{RIS} can be theoretically considered as continuous phase systems, this assumption is not practically feasible due to hardware limitations. Therefore, we consider discrete-phase \acp{RIS} with \ac{GAT} channel estimation. Error rate performance is investigated up to quantized 3-bits \acp{RIS} with the variable number of elements. First, we assume a basic \ac{RIS} design which enable phase shifts, $\phi_n$, in $[-\pi, \pi)$. The step size in phase shift set supported by \ac{RIS} is determined by the number of quantization levels as $\frac{2\pi}{2^{N_{\text{bit}}}}$, where $N_{\text{bit}}$ stands for the number of bits. It can be heuristically said that the system performance improves if the number of distinct discrete phase shifts that \ac{RIS} can provide increases, or in other words, converges to the continuous phase case. \FGR{fig:ber_discrete} gives the \ac{BER} results for the quantization levels of $2$, $4$, and $8$. The simulation results denote that increase in the number of quantization levels for phase shifts reduce\textcolor{black}{s} the error probability. As seen in~\FGR{fig:ber_discrete}, 3-bit \ac{RIS} designs almost approach the error probabilities that the optimum design can achieve. 1-bit designs can only show the same \ac{BER} performance at higher \ac{SNR} values compared to 2-bit designs. In other words, 2-bit designs are much more energy-efficient. In addition, considering the cost of \acp{RIS} that support\textcolor{black}{s} more than 3-bit quantization levels, it is considered feasible to use 3-bit designs for energy-efficient satellite \ac{IoT} communication systems.

\begin{figure}[!t]
    \centering
    \includegraphics[width=0.9\linewidth, page = 4]{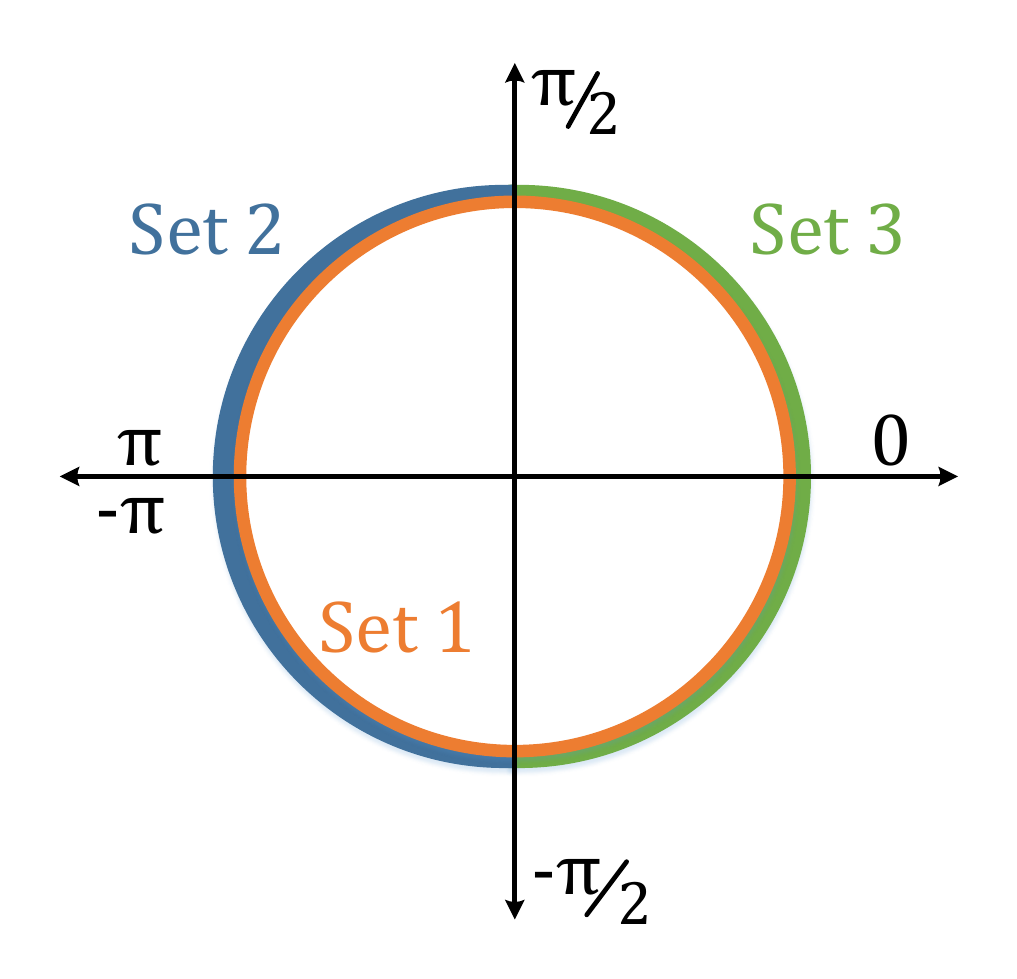}
    \caption{The illustration for the phase sets on the unit circle.}
    \label{fig:discrete_phase_set}
\end{figure}

The phase shifts might not be  uniformly distributed in $[-\pi, \pi)$ because of the limitations in intelligent metasurface designs. For example, the phase shift span is proportional to the square-root of the ratio between the effective capacitance and the effective inductance\cite{liu2019intelligent}. Because the varactor diode reaches the saturation level (i.e., constant capacitance value) although the control voltage increases, all phase shift values may not be supported by the intelligent surface. For instance, the surface in~\cite{tang2019wireless_2} is able to shift the phase of an incident wave by up to $250$ degrees. Hence, we investigate the error rate performance versus the different phase shift sets, namely different \ac{RIS} designs. We evaluate three sets: Set 1 has been already introduced above. Set 2 and Set 3 only consist of the left-hand side and right-hand side of the phase circle, respectively. The phase shift sets are summarized in~\TAB{tab:phase_sets} and illustrated in~\FGR{fig:discrete_phase_set}. Besides, the discrete phase shifts are employed in this analysis. The phase shift, $\phi_n$, is determined as the closest phase value in the set to the phase of the channel coefficient estimated by \ac{GAT} as follows:
\begin{equation}
    \phi_n = S\{\phi_k : k = \underset{s}{\mathrm{argmin}}(|\phi_{s}-\angle(\hat{h}_n \hat{g}_n)|)\},\, s = 1, \cdots, 2^{N_{bit}},
\end{equation} 
where $S$ is the phase shift set including discrete phase shifts, $\phi_{s}$. $\angle(\cdot)$ stands for the angle operator. Additionally, $h_n$ and $g_n$ denote the estimated channel coefficients regarding the $n$-th elements of \ac{RIS}. \FGR{fig:ber_discrete_phase_set} \textcolor{black}{shows} the \ac{BER} performance of \ac{RIS}-assisted satellite \ac{IoT} system for the different phase shift sets with 2- and 3-bits. Set 1 and Set 2 show almost similar \ac{BER}; however,  Set 1 results in slightly better error performance. Moreover, \ac{BER} performance improves when the quantization level increases in the first two sets while increasing the number of bits in Set 3 surprisingly worsens the performance. To explain this, it is necessary to look closely at the channel model. \FGR{fig:angle_histogram} shows the phase histogram of the actual cascaded channel and the phase histogram of the estimated channel coefficients, respectively. As seen, the phase of the cascaded channel is concentrated around -$\pi$ and $\pi$. Likewise, since the channel estimation performance is high, the information about the estimated channels is parallel to the actual channel. These histograms show why Set 1 and Set 2 both performed similarly and have higher performance compared to Set 3. Since the working principle of \ac{RIS} is to make the \ac{SNR} maximum by eliminating the phase information of the cascaded channel, \ac{RIS} must support phase shifts in a way that eliminates the phases of the cascaded channel. Set 1 and Set 2 support the phase shifts of the channels to omit the phases concentrated around -$\pi$ and $\pi$. However, since Set 3 consists of phase shifts between -$\pi/2$ and $\pi/2$, it cannot completely exclude actual phases around $\mp\pi$. In addition, increasing the quantization level also allows correcting the actual phases different from $\mp\pi$ in Set 1 and Set 2. The increase in the number of levels enables the generation of new phase values between -$\pi/2$ and $\pi/2$ in Set 3 and creates phase shifts \textcolor{black}{farther} away from $\mp\pi$. Therefore, the phase information cannot be adequately corrected.
\begin{table}[!t]
\centering
\caption{The phase sets with different spans between -$\pi$ and $\pi$.}
\begin{tabular}{ccc}
\toprule \toprule
\multicolumn{1}{l}{\textbf{Phase Set}} & \multicolumn{1}{c}{\textbf{Phase Interval}} & \multicolumn{1}{c}{\textbf{Step Size}} \\  \midrule
\multicolumn{1}{c}{Set 1}       & $[-\pi, \pi)$     & $\frac{2\pi}{2^{N_{\text{bit}}}}$    \\
\multicolumn{1}{c}{Set 2}       & $(-\pi, -\frac{\pi}{2}] \cup [\frac{\pi}{2}, \pi)$     &  $\frac{\pi}{2^{N_{\text{bit}}}}$    \\
\multicolumn{1}{c}{Set 3}       & $[-\frac{\pi}{2}, \frac{\pi}{2}]$     & $\frac{\pi}{2^{N_{\text{bit}}}}$    \\
\bottomrule \bottomrule
\end{tabular}
\label{tab:phase_sets}
\end{table}

\begin{figure}[!t]
    \centering
    \includegraphics[width=\linewidth]{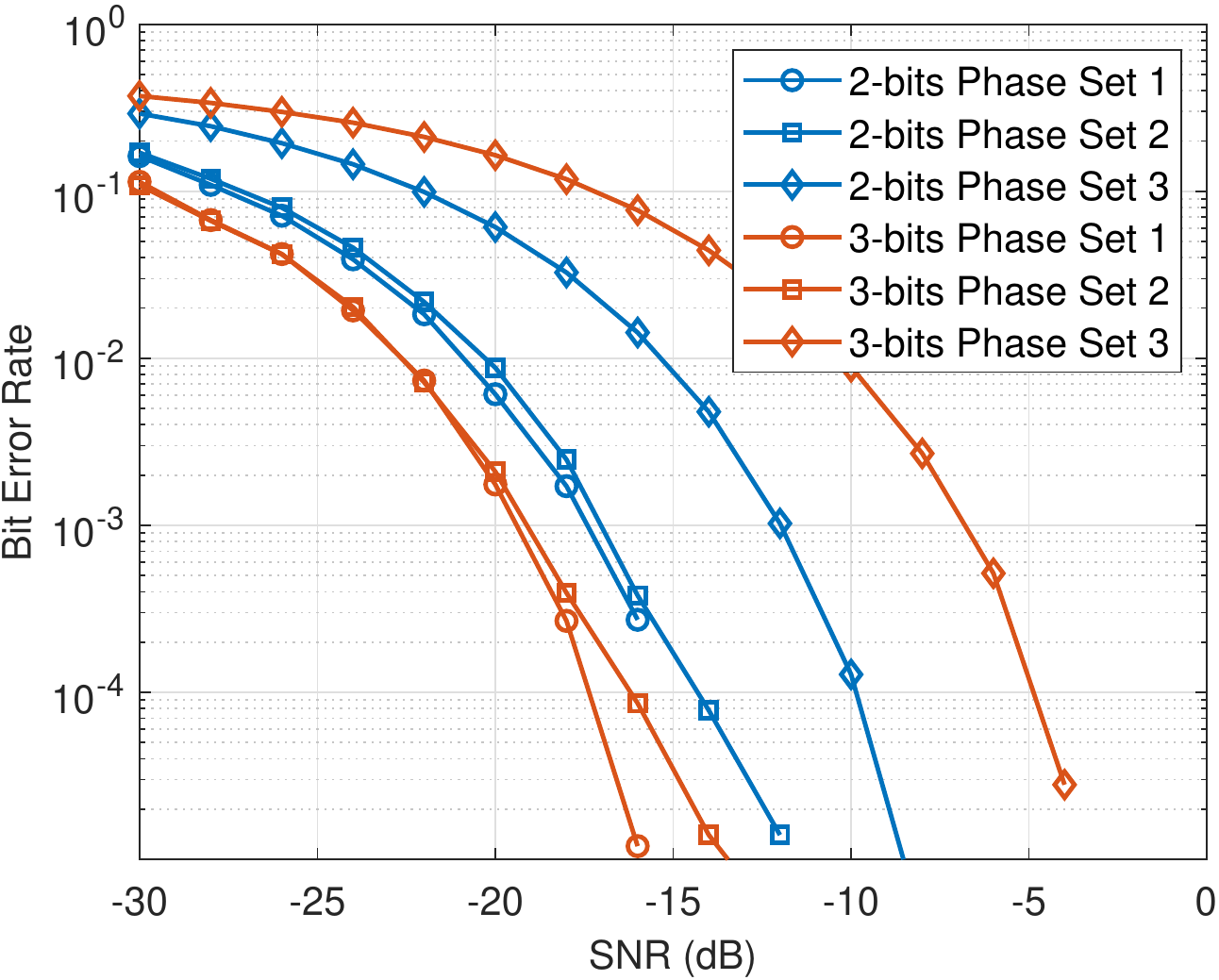}
    \caption{\ac{BER} performance of discrete phase $32$-elements RIS-assisted satellite \ac{IoT} communications regarding different phase sets.}
    \label{fig:ber_discrete_phase_set}
\end{figure}


\begin{figure}[!t]
    \centering
    \includegraphics[width=\linewidth]{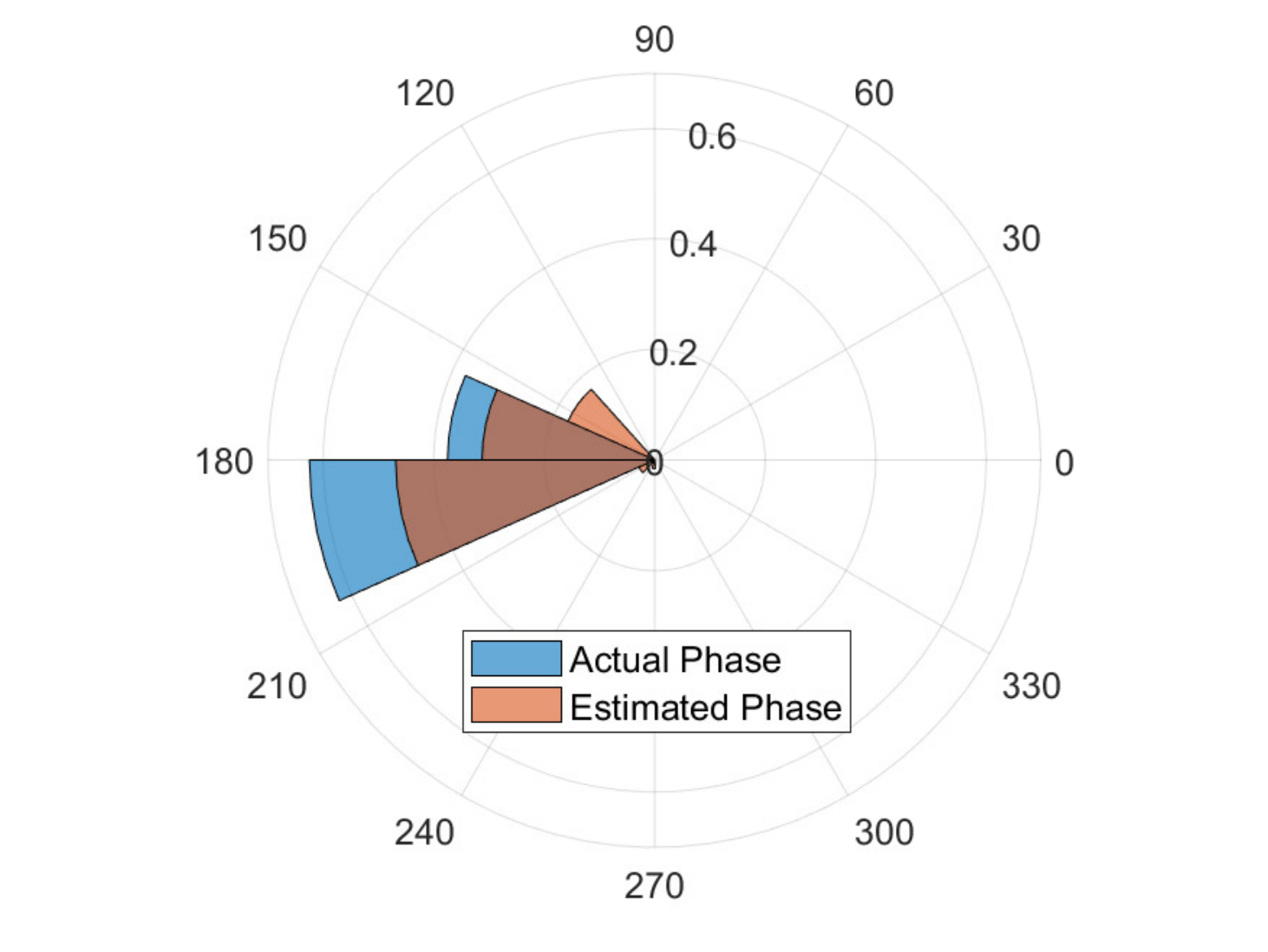}
    \caption{\textcolor{black}{The histogram for the phase of actual and estimated channel coefficients.}}
    \label{fig:angle_histogram}
\end{figure}

\section{Research Directions and Concluding Remarks}\label{sec:conclusion}
In this study, \ac{GAT} based channel estimation is presented for \ac{RIS}-assisted communications and it is shown that high performance is achieved. The proposed channel estimation method based on GAT is able to learn inductively; therefore, it can provide high performance under changing conditions that are not included in the training phase.

To improve \ac{DtS} \ac{IoT} systems, \ac{RIS} with \ac{GAT} is integrated to the system architecture. By doing so, it is demonstrated that the error probability is achieved at lower \ac{SNR} compared to conventional methods. 
\ac{BER} performance under various RIS configurations including discrete and piecewise phase sets is numerically investigated. It is shown that 2-bit resolution can almost perform as well as ideal RISs with a continuous phase shift. In addition, the simulation results show that the channel model is an important design parameter \textcolor{black}{for} RISs. This study provides a comprehensive investigation of GAT-based channel estimation and performance analysis on RIS-assisted \acp{DtS}; however, the impact of Doppler shift to GAT performance due to the motion of LEO satellites requires more investigation. \textcolor{black}{It is shown in this study that the proposed GAT-based estimator performs better with much shorter pilot signals compared to the state-of-art methods in the literature. It should be noted \textcolor{black}{that} although very low pilot overhead is required with the proposed method, there may be performance losses in estimating the instantaneous CSI due to the movement of LEO satellites. Therefore, in future studies, it is planned to work on performance improvement with the use of average CSI based on the method given in this \textcolor{black}{study}.} Moreover, RIS fabrication and deployment on satellites requires consideration of space conditions and SWaP constraints, so it should be the subject of interdisciplinary research. \textcolor{black}{Another important point is investigating system performance for the downlink signaling since this study covers only performance analysis for uplink. Due to partial channel reciprocity, it is possible to observe some changes in channel characteristics between uplink and downlink. However, as shown in this study, the GAT-based estimator can provide generalizable results and an adjustment factor can be used with respect to satellite elevation angle to estimate the changes in the downlink channel compared to the uplink channel estimated by the GAT. Furthermore, as impairments in RIS design can seriously affect performance, future studies should consider impairments. For instance, as this study considers all-passive element RIS, the saturation effect is not taken into account. However, the saturation effect must be considered for RIS designs including active elements.}

\textcolor{black}{Last but not least, even though this study mainly addresses DtS IoT communication, cooperation between RISs and aggregator gateways which are mainly used in the state-of-the-art satellite IoT applications should be investigated in future studies.}

\balance

\bibliographystyle{IEEEtran}
\bibliography{main}

\balance
\end{document}